\documentclass[prb,twocolumn,showpacs,amsmath,amssymb]{revtex4}

\usepackage{amsmath}
\usepackage{graphicx}
\usepackage[pdftex,dvips]{epsfig}
\newcommand{\be}{\begin{equation}}
\newcommand{\ee}{\end{equation}}
\newcommand{\bma}{\begin{displaymath}}
\newcommand{\ema}{\end{displaymath}}
 
\begin{document}

\title{Localization of particles in harmonic confinement: Effect of
  the interparticle interaction}

\author{J.-P. Nikkarila and M. Manninen}

\affiliation{\sl NanoScience Center, Department of Physics,
FIN-40014 University of Jyv\"askyl\"a, Finland}

 
\date{\today}

\begin{abstract} 

We study the localization of particles rotating 
in a two-dimensional harmonic potential by solving 
their rotational spectrum using many-particle quantum
mechanics and comparing the result to that obtained with
quantizing the rigid rotation and vibrational modes of 
localized particles. We show that for a small number of
particles the localization is similar for bosons and
fermions. Moreover, independent of the range of the
interaction the quantum mechanical spectrum 
at large angular momenta can be
understood by vibrational modes of localized particles.

\end{abstract}

\maketitle

\section{introduction}

Interacting particles rotating in a harmonic confinement has 
become an experimentally realizable quantum system which has
attracted plenty of attention during the last decade. 
The experiments in semiconductor quantum dots in the presence
of a magnetic field showed Hund's rule\cite{tarucha1996}
and properties related to the 
quantum Hall liquids\cite{mceuen1991,oosterkamp1999}.
In increasing magnetic field the electrons, 
polarized by the field, 
first form so-called maximum density
droplet\cite{macdonald1993},
which is the finite electron number counterpart of the 
integer quantum Hall effect, and then are expected to form
fractional quantum Hall liquid\cite{laughlin1983,maksym1990,wojs1997}
and eventually get localized to a Wigner crystal\cite{maksym1996}.

The weakly interacting boson and fermion gases in 
atomic traps are other systems where particles move 
in a nearly harmonic trap. The traps can be three-dimensional
or quasi-two-dimensional. Rotating laser field has been used to 
put the atom cloud in rotational motion which has lead to
observation of vortices in both boson and fermion
systems\cite{madison2000,abo2001,zwierlein2005}.

The experiments have inspired numerous theoretical studies 
on the properties of rotational quantum states of 
particles in a harmonic confinement. For quantum dots 
these studies have revealed several kinds of internal
symmetry breakings of the system: Spin-density waves\cite{koskinen1997},
broken symmetry edge\cite{reimann1999}, 
vortex formation\cite{saarikoski2004,manninen2005}
and electron
localization\cite{muller1996,yannouleas1999,reimann2000}. 
In the case of boson
and fermion condensates the theoretical research has 
concentrated on quite small angular momenta and has mainly been 
interested in vortex 
formation\cite{butts1999,mottelson1999,bertsch1999,toreblad2004,reimann2006a},
although localization by rotation has also 
been studied\cite{manninen2001,romanovsky2006,reimann2006}. 
It has been observed that both the vortex formation\cite{toreblad2004} 
and the localization\cite{reimann2006} are qualitatively similar 
for both types of particles, and that the type of the
interparticle interaction does not seem to play an 
important role.

In this paper we will study in detail the limit
of the particle localization to a 'Wigner molecule'
where the many-particle spectrum can be determined 
by combining classical vibrational modes to the 
rigid rotation of the 'molecule'. In the case of electrons
this limit was studied in detail by
Maksym\cite{maksym1996,maksym2000} and later 
by us\cite{nikkarila2007}.
Here we will extend this work to bosonic systems and
also study the effect of the range of the interparticle interaction.

The paper is organized as follows. In Section II we introduce
the models and discuss an exactly solvable model of 
harmonic interparticle interaction. In Section III 
we present results, first for seven particles interacting
with Coulomb interaction and then study the effect
of the range of the interaction using as an example
four particles interacting with a Gaussian interaction.
The conclusions are given in Section IV.

\section{Theoretical models}

\subsection{Quantum mechanics in the lowest Landau level}

We assume a generic model of spinless fermions or bosons
in a two-dimensional harmonic
potential. The Hamiltonian is on the form
\be
H=-\frac{\hbar^2}{2m}\sum_{i}^N \nabla_i^2 
+\sum_i^N \frac{1}{2}m\omega_0^2 r_i^2
+\sum_{i<j}^N v(\vert {\bf r}_i-{\bf r}_j\vert),
\label{hamiltonian}
\ee
where $N$ is the number of particles and $m$ the electron mass
and $\omega_0$ the oscillation frequency of the confining potential.
The variable ${\bf r}=(x,y)$ is a two-dimensional position vector. 
The interparticle potential $v$ is either the Coulomb interaction
or a Gaussian repulsion:
\be
v(r)=\frac{e^2}{4\pi\epsilon_0r}\quad {\rm or}\quad 
v(r)=\frac{1}{\pi\sigma^2}e^{-r^2/\sigma^2}.
\label{spstate}
\ee
We solve the many-particle problem for a fixed angular momentum
using the single particle basis of the lowest Landau level (LLL):
\bma
\psi_\ell(r,\phi)=A_\ell r^\ell e^{-m\omega_0r^2/2\hbar}e^{i\ell\phi},
\ema
where $\ell$ is the single-particle angular momentum
and $A_\ell$ a normalization factor. 
The restriction to the LLL is a good approximation when
total angular momentum $L$ is relatively large,
$L\gtrsim N(N-1)$
\cite{manninen2001,jain2005a},
corresponding to the filling factor of about $\nu\lesssim 1/2$ 
of the LLL.
At this point it is convenient to stress that,
when we consider electrons, we do not explicitly 
include magnetic field in our calculations.
Assuming complete polarization, 
the only effect of the magnetic field has, is to increase
the total angular momentum.

We present our results in terms of the total angular momentum,
which can be related to the filling fraction of the LLL
by $\nu\approx N(N-1)/2L$. Angular momenta
$L=qN(N-1)/2$ then correspond to the fractional 
quantum Hall liquid with $\nu=1/q$, where $q$ is the
exponent of the Laughlin wave function\cite{laughlin1983}
(Eq. (4) below).

The restriction of the basis to the LLL gives
some important exact results about the 
properties of the quantum states. These can be seen by
writing the Hamiltonian (1) in the 
occupation number formalism as
\be
\hat H=\hat H_0+\lambda\hat V=\hbar\omega_0N+
\sum_\ell \hbar\omega_0\ell c_\ell^\dagger c_\ell
+\lambda\sum_{ijkl} v_{ijkl}c_i^\dagger c_j^\dagger c_lc_k,
\ee
where $c_\ell^\dagger$ creates a particle in the LLL.
The states are written as $\vert n_0n_1n_2\cdots\rangle$ where 
the subsrcipt refers to the single particle angular momentum
$\ell$.

We notice immediately that for a fixed total angular momentum
$L=\sum_\ell n_\ell$ the diagonal term $\hat H_0$ of the Hamiltonian 
gives a constant $\hbar\omega_0(L+N)$, i.e. it is independent of the 
configuration. The diagonalization of the Hamiltonian thus
equals to diagonalization of the interaction part $\hat V$ of
the Hamiltonian, which gives directly the
{\it interaction energy} $\Delta E=\langle \Psi\vert\hat V\vert\Psi\rangle$. 
Moreover, this means that the structure of the quantum states
are independent of $\omega_0$ and the coupling constant $\lambda$,
which have only the roles of giving the distance and energy scales.

In the case of bosons the state with
$L=0$ is the Bose condensate where $n_0=N$ and all the other
states are empty. for example, for seven bosons the state is
$\vert 7000\cdots\rangle$. In the case of fermions
the lowest possible angular momentum is $L_{\rm MDD}=N(N-1)/2$
and the state is the so-called maximum density droplet\cite{macdonald1993}
$n_i=1$ for $i<N$ and this state for seven fermions
is $\vert 111111100\cdots\rangle$. 
The only possible state at
for seven bosons at $L=1$ is $\vert 61000\cdots\rangle$ and the
corresponding state for seven fermions at $L=L_{\rm MDD}+1$ is
$\vert 11111101000\cdots\rangle$.

It follows that the interaction energy of the $L=0$ (or MDD) state
equals to that of the $L=1$ (or $L_{\rm MDD}+1$) state, 
for any two-body interaction. This is due to the fact that  
the $L=1$ state is the center of mass excitation of the 
$L=0$ state. 
In the LLL the two lowest angular momentum states 
for both bosons and fermions are thus 
completely fixed by the basis and, consequently, are 
independent of the interparticle interaction.

In the case of the delta function interaction, the spinless 
fermions do not feel the interaction due to the Pauli exclusion 
principle. The MDD can be written in the form of the Laughlin
wave function (without normalization) as
\be
\Psi_q=\prod_{i<j}^N(z_i-z_j)^q
e^{\sum m\omega_0^2 \vert z_k\vert^2/2\hbar},
\label{laughlin}
\ee
where $z=x+iy$, and the exponent $q=1$ corresponding to the 
filling factor $\nu=1/q=1$. Since any wave function for higher
angular momenta can be written as a symmetric polynomial times
the above, the effect of the delta function interaction disappears.
In the case of bosons the delta function interaction has an effect
up to the angular momentum $L=N(N-1)$ when a symmetric
wave function of the type
of Eq. (\ref{laughlin}) can be reached, with $q=2$.

It turns out that also for long range interactions,
even for the Coulomb interaction, the Laughlin type wave function
becomes rather accurate for $q=3$ and 5 for fermions and 
for $q=2$ and 4 for bosons, while at even higher values of
angular momenta the particles localize to Wigner molecules which 
rigidly rotate in the harmonic well.

While the Laughlin state gives good estimates for the wave
function only for particular angular momenta, the composite fermion
picture of Jain\cite{jain1989,jain1990,jeon2004} 
gives estimates for all angular momenta.
It is important to note that both Laughlin and Jain wave functions
are derived from very general arguments and are completely 
independent of the interparticle interactions.

The repulsive harmonic interaction
\be
v(\vert {\bf r}_i-{\bf r}_j\vert)=-\lambda \vert {\bf r}_i-{\bf r}_j\vert^2
\ee
is exactly solvable\cite{lawson1980,johnson1992,ruuska2005}.
The resulting interaction energy of the lowest state
as a function of the angular momentum
is for bosons \cite{ruuska2005}
\be
\Delta E(L)=-\frac{\lambda}{\omega_0}N(N-1)-
\frac{\lambda}{\omega_0}NL\theta(L-2)
\label{harminten}
\ee
where $\theta$ is the step function.
Due to the symmetry of the interaction the degeneracy of each
state is large. Note that the interaction energy in this model 
decreases linearly with $L$.

\subsection{Classical particles}

Classical electrons in a 2D harmonic potential arrange in consecutive 
circles i.e. they form Wigner molecules\cite{bolton1993,bedanov1994}.
The molecule can have angular momentum due to the center of mass 
rotation or due to the rigid rotation of the molecule. In addition
the molecule can have internal vibrational motion.
The internal vibrational modes can then be found 
by solving the dynamical 
matrix. We are interested in the excitations of a
rotating system and therefore we have to solve the
internal vibrations in a rotating frame of reference.
For quantum dots this was first done by 
Maksym\cite{maksym1995,maksym1996} 
who used the Eckart rotating frame.
We use a more straigthforward method where the Newton equations
of motion are determined in a standard rotating frame,
taking into account the Coriolis force\cite{goldstein1969}.
We fix the rotation frequency $\omega_r$ which determines
the total angular momentum $L$ of the rotating molecule.

The equilibrium positions of the particles depend naturally
on the angular frequency and total angular momentum
$L=I \omega_r$, where $I$ is the moment of inertia
$I=m \sum r_i ^2$. The classical energy 
\be 
E_{\rm cl}^0(L)=\frac{1}{2}m\omega_0\sum_i^N r_i^2
+\sum_{i<j} v( {\bf r}_i-{\bf r}_j )
+\frac{L^2}{2I}
\label{Eclassical}
\ee
is minimized in order to solve the equilibrium configuration.
The interparticle interaction $v( \vert{\bf r}_i-{\bf r}_j \vert)$ 
can be either Coulombic
or Gaussian just like when solving the energy spectra of corresponding 
quantum mechanical systems.  

In this article we report
results for up to seven bosons and fermions. 
The eigenfrequencies can be solved analytically (or
numerically) from the equations of motion in the rotating frame:
The Newton equations are linearized around around
the equilibrium positions of the particles and
the vibrational modes are obtained with matrix algebra. 
When the classical vibrational frequencies are solved
we quantize the energies with 
canonical quantization\cite{ashcroft1976} and
get an estimate for the total quantum mechanical energy, which
in general can be written as
\be
E_{\rm cl}(L)=E_{\rm cl}^0(L)+\sum_k \hbar\omega_k(L)(n_k+\frac{1}{2})+
\hbar\omega_0(n_{\rm CM}+1),
\label{Ecltot}
\ee
where $\omega_k(L)$ is the vibrational frequency determined in the 
rotating frame with angular momentum $L$, 
$n_k=0,~1,~2,\cdots$, and the last term
corresponds to the center of mass excitations.
The symmetry or antisymmetry requirement of the total quantum
state dictates which vibrational states are allowed for a
given $L$. For this reason we have to determine the 
symmetry properties of the vibrational modes and use group 
theory\cite{tinkham1964,ruan1995,maksym1996,koskinen2001,koskinen2002}
to resolve the allowed states.
Using the notations of Maksym\cite{maksym1996}
we have the following requirements for fermions
\begin{displaymath}
L+\sum_i n_i k_{i} = 
\bigg\{ \begin{array}{ll}
0 \text{ mod } m  & \text{ for } m= \text{odd} \\
m/2 \text{ mod } m   & \text{ for } m= \text{even}.
\end{array}
\end{displaymath}
The corresponding requirement for both even and odd number of bosons is
\begin{displaymath}
L+\sum_i n_i k_{i} = 
0 \text{ mod } m .
\end{displaymath}
The symbols $n_i$ refer to the number of excitations
of mode $\omega_i$ and the number $ k_i$
to the symmetry property of the same mode
(rotation by $2\pi k/m$ is presented by $\exp(i2\pi k/m)$)\cite{maksym1996}.

The classical system with a repulsive harmonic interaction
can also be solved analytically. For the lowest energy state
as a function of the angular momentum we get
\be
E(L)=L\sqrt{\omega_0^2-2N\lambda}.
\ee
Quantizing this and taking into account the zero point 
energy, the interaction energy will be
\be
\begin{split}\Delta E(L)=&N\omega_0\sqrt{1-\frac{2(N-1)\lambda}{\omega_0^2}}\\
&+L\omega_0\sqrt{1-\frac{2N\lambda}{\omega_0^2}}-(N+L)\omega_0,
\end{split}
\ee
which for small $\lambda$ gives the result of Eq. (\ref{harminten}).
Notice that for $\omega_0^2<2N\lambda$ the harmonic potential
is not strong enough for confining the particles.

We should also notice that in the case of the delta function 
interaction the classical system does not have any vibrational
modes, but the particles behave like noninteracting particles
(the collision probability is zero).

\section{Results}

\subsection{Seven particles with Coulomb interaction}

We have considered four and seven bosons and fermions
with different interparticle interactions. 
We remind the reader that in the LLL the results are independent
of $\omega_0$, apart from the energy scale.
It is then convenient to present the results in atomic units 
($m=\hbar=e=a_0=1$). We use $\omega_0=1$ for the Gaussian interaction
and $\omega_0=1/2$ for the Coulomb interaction.

The classical model has been studied for electrons in quantum
dots by Maksym et al\cite{maksym1996,maksym2000}, 
Matulis and Anisimovas\cite{matulis2005} and 
by us\cite{nikkarila2007}. We will first demonstrate that 
the same model works also for bosons interacting with the
Coulomb interaction. Since the earlier computations have been 
restricted to six or less electrons we do this comparison for
seven particles.

The classical problem of seven charged particles 
in a 2D harmonic potential can be
solved analytically by first determining the classical minimum
energy \eqref{Eclassical} at the equilibrium 
in the rotating frame and then solving the dynamical matrix.
The lowest energy classical configuration
is a hexagon of six particles surrounding the seventh 
particle at the origin.
This geometry naturally has a six-fold rotation axis.
The other stable classical geometry is a seven-fold ring.
However, its energy is so much higher that it does not 
contribute to the low-energy rotation-vibration spectrum.

\begin{figure}[h]
\includegraphics[width=0.95\columnwidth]{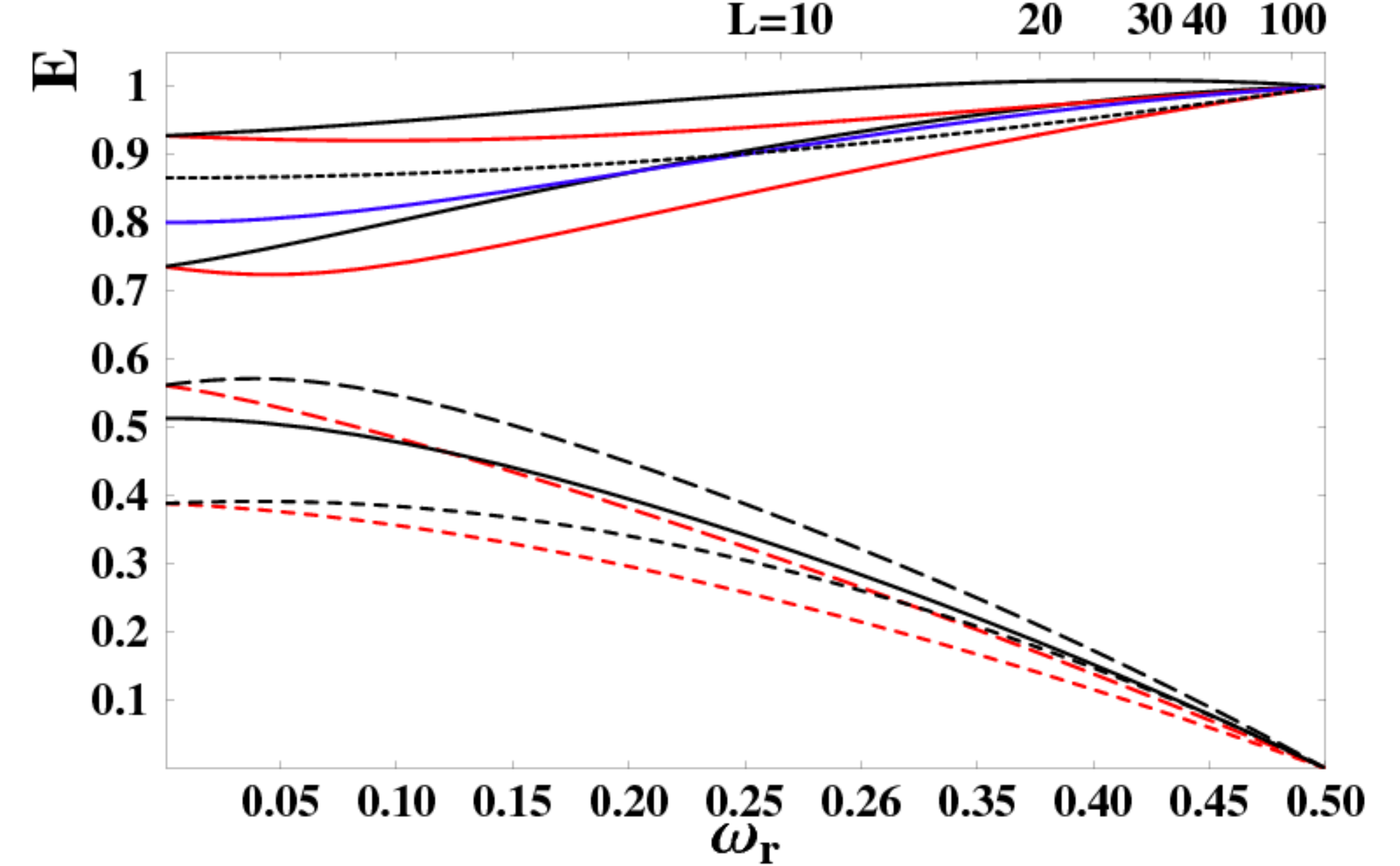} \vspace{2pt} \\
\includegraphics[width=0.3\columnwidth]{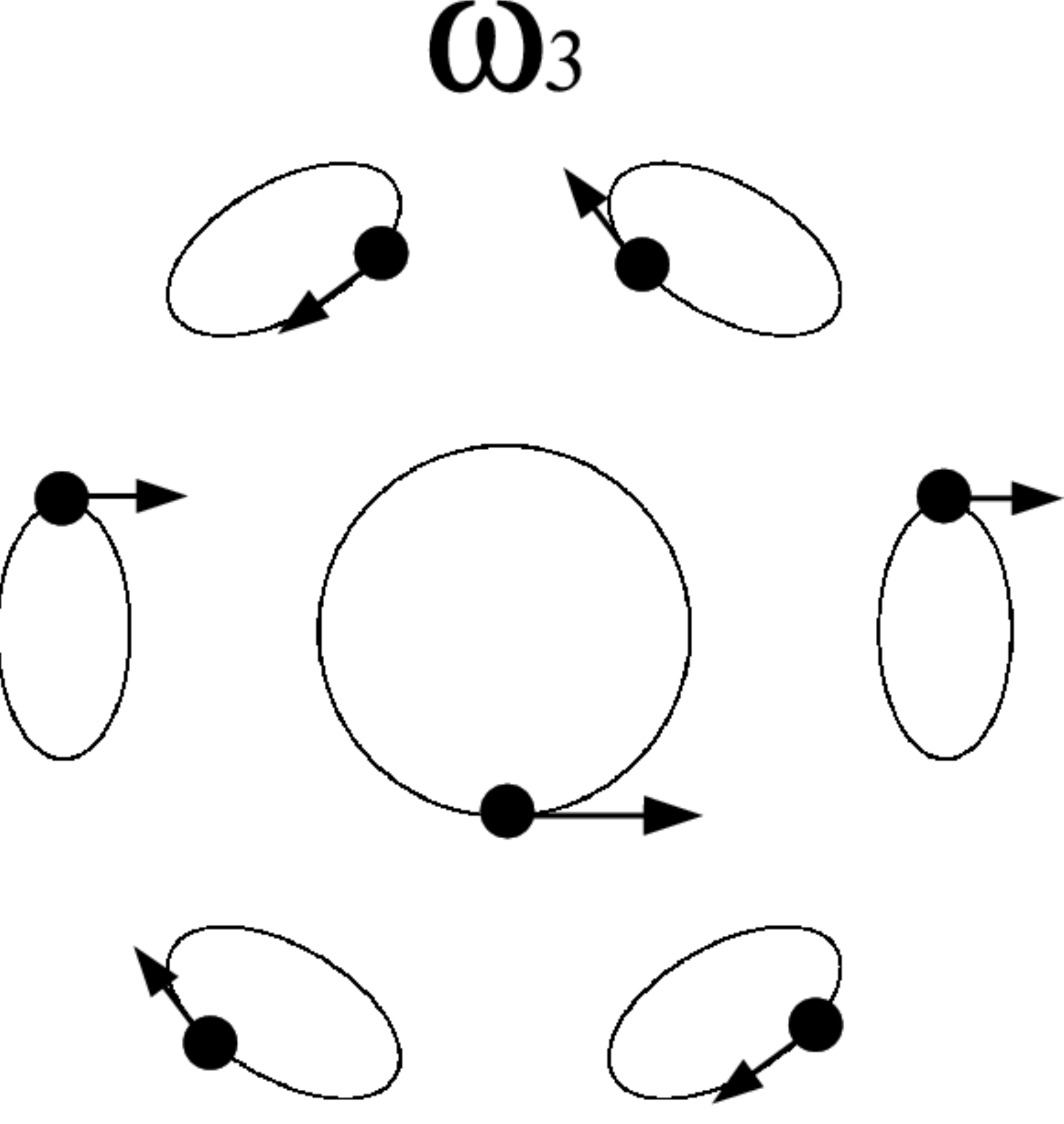}\hspace{2pt}  
\includegraphics[width=0.3\columnwidth]{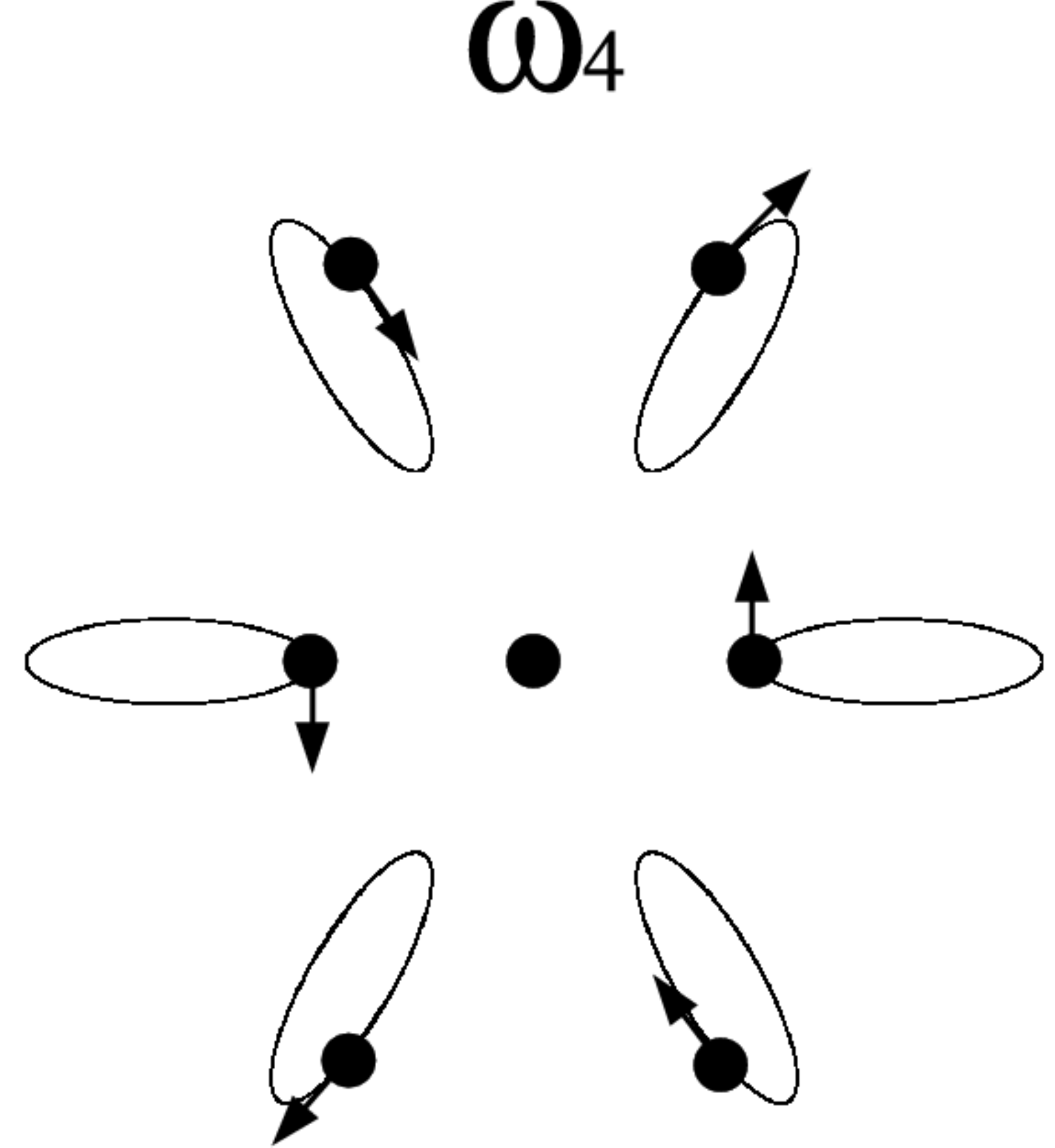}\hspace{2pt} 
\includegraphics[width=0.3\columnwidth]{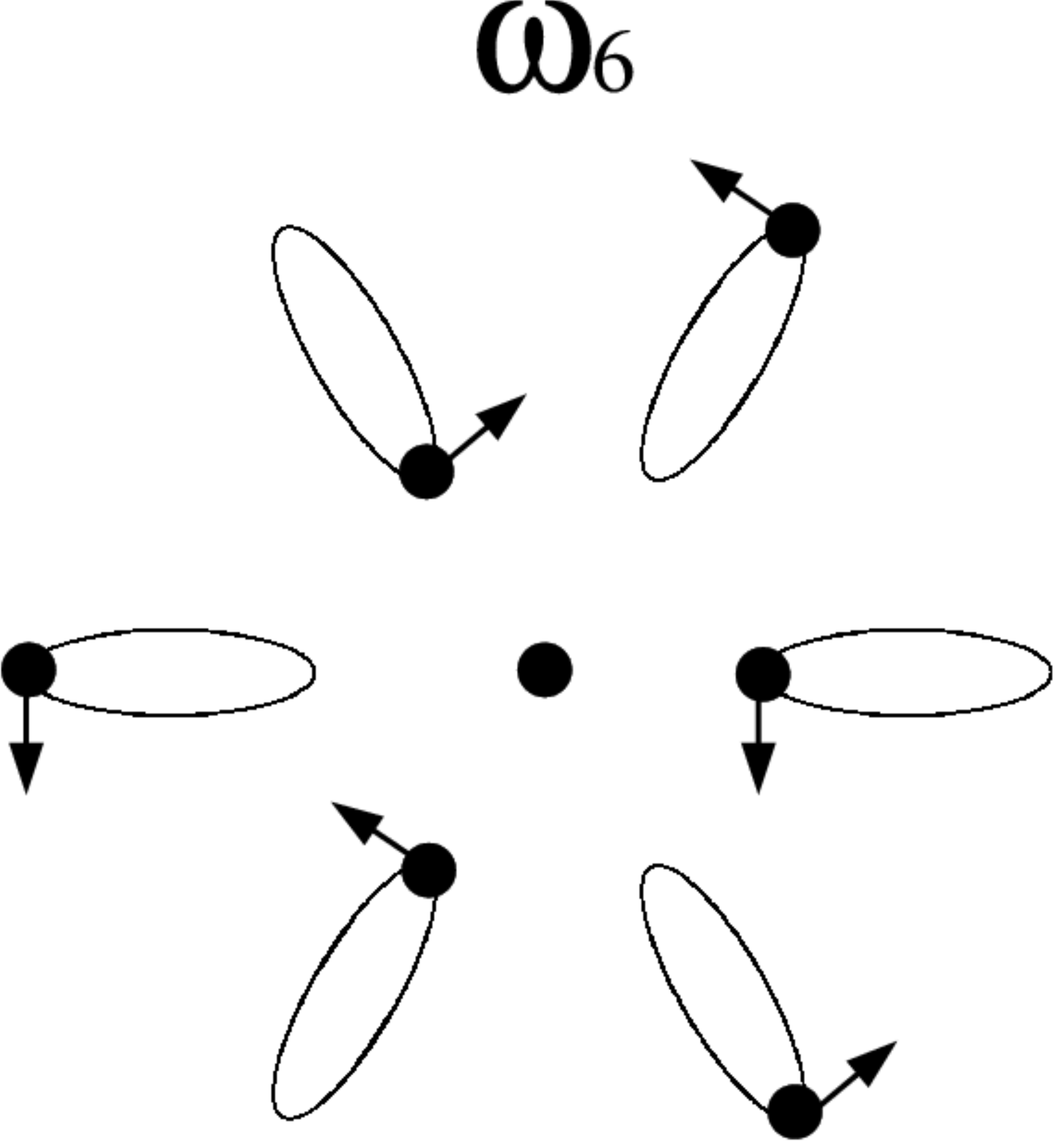}
\caption{Upper panel: Classical vibrational frequencies as a function
of the angular velocity of the rotating Wigner molecule
with seven charged particles. The breathing mode is shown as a dotted line.
The dashed and solid lines show other modes for seven electrons
(in atomic units with $\omega_0=0.5$).
Lower panel: Modes of the low energy vibrations (see text). 
Mode $\omega_5$ is similar to the
mode $\omega_3$ (and degenerate at $\omega_r=0$) but the motion of the
particles is opposite to that of mode $\omega_3$.
In mode $\omega_4$ the particle at the center does not move. 
Mode $\omega_7$ is similar to mode $\omega_4$ (degenerate
at $\omega_r=0$) but the particles move opposite direction.
Note that in mode $\omega_6$ the orbits of the particles are the
same as in modes $\omega_4$ and $\omega_7$ but the particles are
at different phases.
}
\label{clseven}
\end{figure}

The classical vibrational frequencies 
(except the center of mass mode) for 
7 particles are shown in Fig. \ref{clseven}.
The eigenfrequencies are shown 
as a function of the angular velocity of the rotation $\omega_r$.
The angular momentum of the system increases rapidly and diverges as 
$\omega_r\rightarrow \omega_0$ (0.5 in this case). Some values of 
the angular momenta are shown in the figure.
We notice that all frequencies approach to either $0$ or $2\omega_0$.
One of the internal vibration energy is on the form 
$\omega_{\rm bm}=\sqrt{3\omega_0^2+\omega_r^2}$, and it is shown in the
figure as a dotted line.
This is a well known result for electrons forming 
one single ring\cite{schweigert1995,maksym1996}
and it is called the breathing mode.
The most interesting modes are the ones that approach $0$.
We have labeled these low-energy modes as $\omega_3$ (lowest mode,
plotted as red dashed line, $\omega_3 \approx 0.4$ when $\omega_r =0$),
$\omega_5$ (dashed line, is degenerate with $\omega_3$ when $\omega_r =0$),
$\omega_6$ (solid line), $\omega_4$ (red dashed line, $\omega_4
\approx 0.55$ when $\omega_r =0$
and as $\omega_7$ (the mode that is degenerate with  $\omega_4$ when
$\omega_r =0$ and $\omega_7 > \omega_4$  elsewhere).
We denote the center of mass vibrations as  $\omega_1$ and $\omega_2$.
The symmetry properties (i.e $k$ values) are
determined as in \cite{maksym1996} and they are
for the low energy vibrational modes of seven electrons
as follows: $k_1=5$, $k_3=1$, $k_4=4$, 
$k_5=5$, $k_6=3$ and $k_7=2$.
We have visualized the electron motion in 
some of modes in the Fig. \ref{clseven}.

\begin{figure}[h]
\includegraphics[width=1.0\columnwidth]{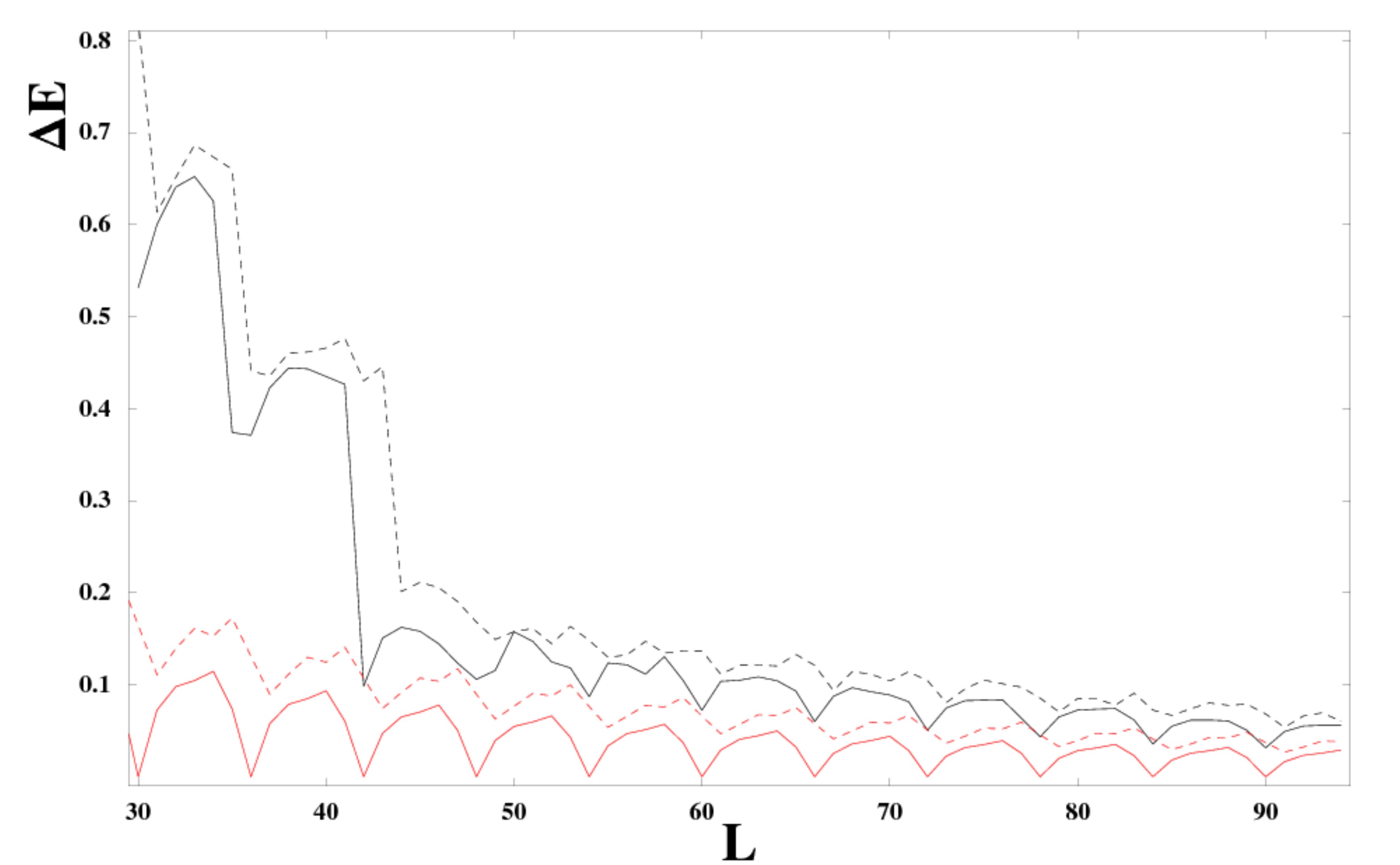}
\includegraphics[width=1.0\columnwidth]{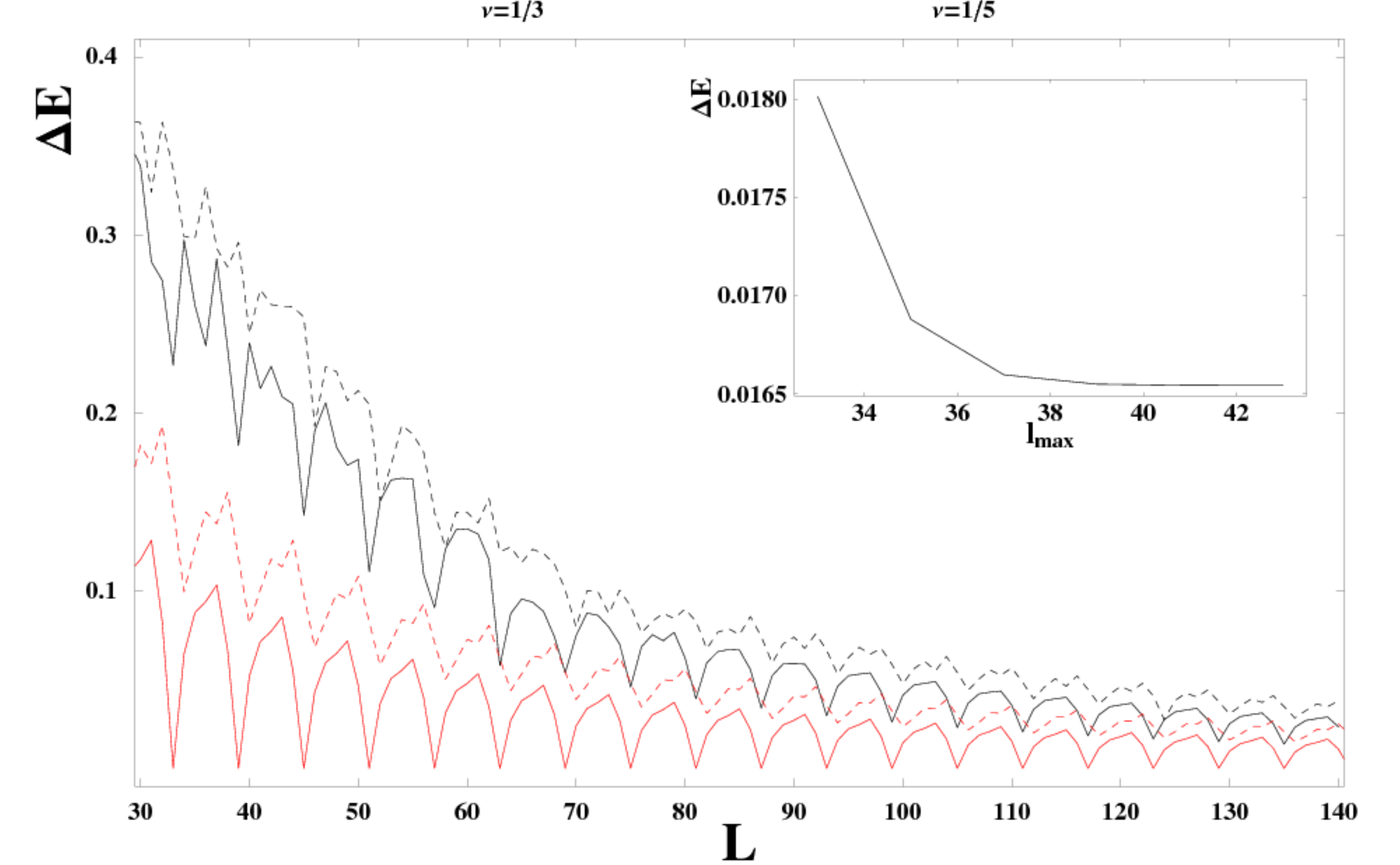}
\caption{Dependence of the ground state energy (solid lines) and the 
first excites state energy (dashed lines) on the total angular
momentum in the case of seven particles interacting
with the Coulomb interaction.
The black lines (red lines) show the energy difference
between the result of the
exact diagonalization (model Hamiltonian (8)) 
and the energy of the rigid rotation 
of the classical Wigner molecule. The upper panel is for bosons
the lower panel for fermions. 
The inset shows the energy convergence as a function of the 
cut-off angular momentum of the LLL single particle basis
in the case of fermions with $L=123$.
}
\label{bf7}
\end{figure}

Once we have the energies and the symmetry properties of the
classical vibrations solved, we need to quantize them 
to get the rotational spectrum of Eq. (8).
Determination of the symmetry properties
means that for fermions we demand that the quantized state is
antisymmetric and for bosons symmetric.

Figure \ref{bf7} demonstrates how the quantum mechanical 
spectrum approaches to that obtained by quantizing the classical
vibrations. The figure shows the lowest (so-called yrast state)
and second lowest energy for each angular momentum. 
The energies are shown as the energy difference from the
classically determined rotational energy, Eq. (8) with
$n_k=n_{\rm CM}=0$. 

The results show the characteristic oscillation as a 
function of the angular momentum. The oscillation can be
understood as a rigid rotation of the 'molecule' of 
localized particles. The symmetry properties allow only 
every sixth angular momentum due to the six-fold symmetry
of the hexagon. For bosons the allowed angular momenta
are $L=6n$ and for fermions $L=6n+3$, where $n$ is an integer.
For angular momenta in between, the purely rotational
state has to be accompanied by a vibrational state, and consequently
the energy is higher.

Figure \ref{bf7} shows that already 
from angular momentum of about 
$L=40$, corresponding to filling factor $\nu=1/2$, 
the model Hamiltonian (8) describes qualitatively well
the oscillations of the energy as a function of the
angular momentum, when compared to the results
of diagonalization of the Hamiltonian (3).
The agreement gets better with 
increasing angular momentum. Naturally, the spectra
for fermions and bosons become then identical, apart of a 
phase shift dictated by the symmetry requirements.

The reason that at high angular momenta the absolute energies of
the exact diagonalization remains higher than those of the 
model Hamiltonian is due to the anharmonicity of the
classical vibration potential: The zero point energy
contribution coming from the high energy modes of 
Fig. 1 causes the energy shift
(i.e. the zero-point energy is underestimated due to the 
harmonic approximation).
The restriction of the basis within the lowest 
Landau level or the cut-off used
for large angular momenta do not play any marked role.
The inset in Fig. 2 shows the dependence of the energy on the cut-off 
single particle angular momentum for $L=123$.
Note the different energy scale and that 
the total energy indeed depends very little on 
the cut-off angular momentum.

\begin{figure}[h]
\includegraphics[width=1.0\columnwidth]{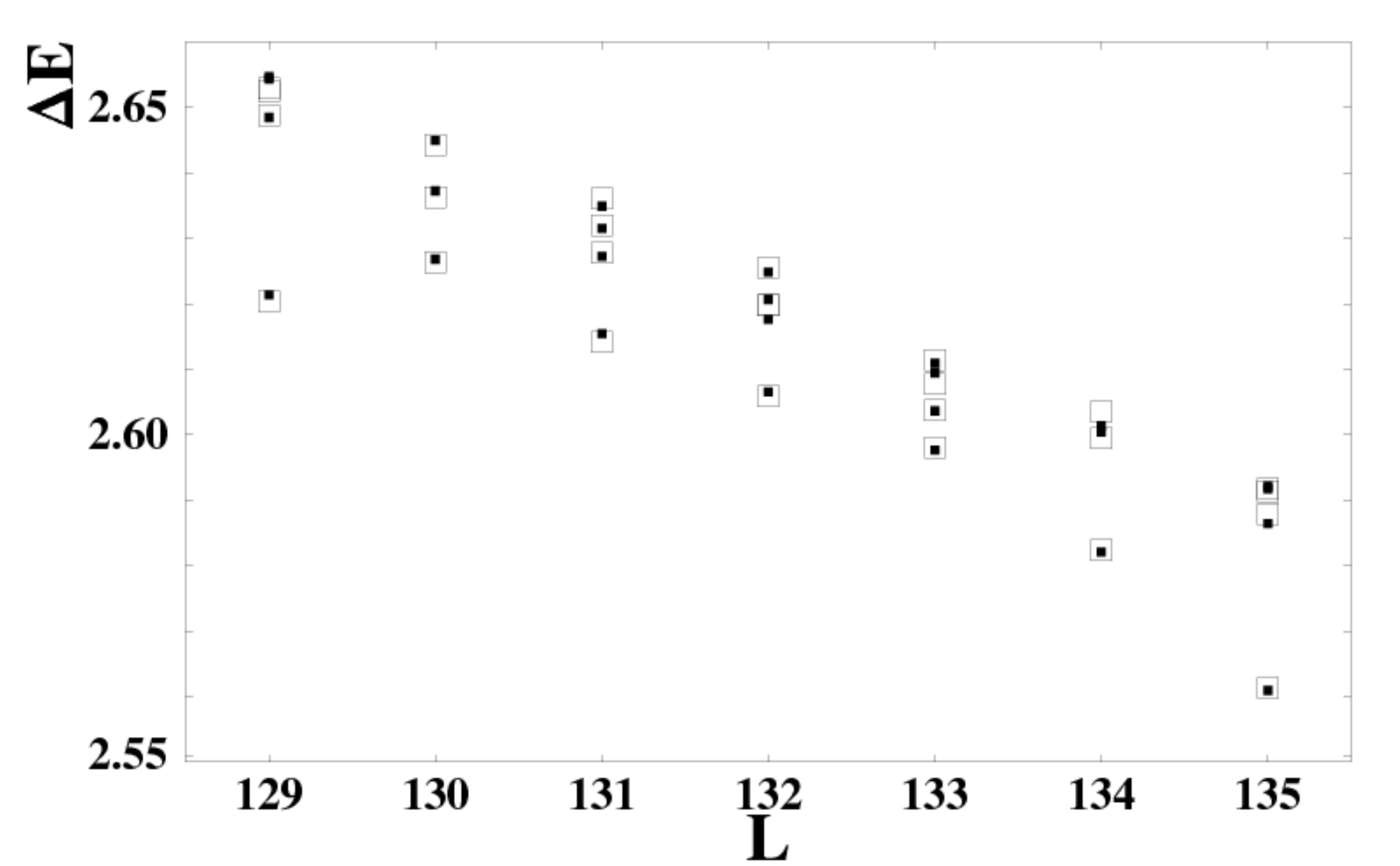}
\caption{Low energy spectrum of seven electrons 
at high angular momenta. Black dots show the 
exact diagonalization results and open squares the 
results of the model Hamiltonian (8), shifted upwards
with a constant 0.014. $\Delta E$ 
is the interaction energy in atomic units
($\omega_0=0.5$). The ground states for $L=129$
and 135 are purely rotational states. The ground states
from $L=130$ to $L=134$ have vibrational modes
$\omega_5$,  $\omega_4$, $\omega_6$, $\omega_7$, $\omega_3$,
respectively.
}
\label{f7}
\end{figure}

Figure \ref{f7} shows in more detail
part of the low energy spectrum for seven
electrons calculated by two different ways, quantum
mechanically and classically. We remind
that in both cases the values are for {\it interaction energy}
i.e $\Delta E =E-(N+L)\hbar\omega_0$, where $E$ is
either the solution from the diagonalization of the quantum 
mechanical Hamiltonian (3) or the classical
energy \eqref{Ecltot}. In Fig. 3 the classical spectrum is
shifted up by a constant energy (0.014 a.u.) to make the comparison
of the structure of the spectra easier. We can see that 
the two spectra agree in all details.
Note also that the difference in the interaction energy
given by the two models, $0.014/2.60\approx 0.5 \%$, 
is very small indeed. 

The oscillatory behavior of the results presented in Figs. \ref{bf7} and
\ref{f7} is not surprising since
similar results were reported by Maksym \cite{maksym1996,maksym2000}
for three, four, five and six electrons in a quantum dot. 
Here we have shown that similar spectrum is obtained for 
fermions and bosons and the approach to the classical
limit is similar in both cases.

\begin{figure}[h]
\includegraphics[width=0.32\columnwidth]{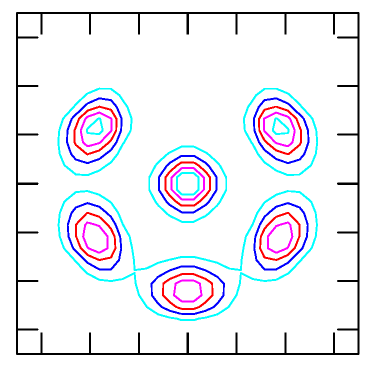}
\includegraphics[width=0.32\columnwidth]{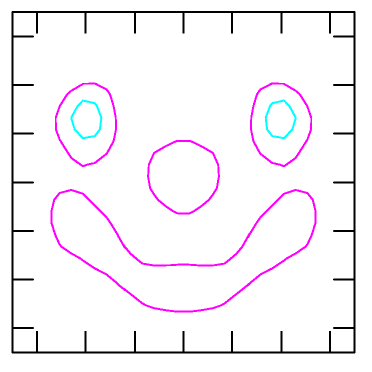}
\includegraphics[width=0.325\columnwidth]{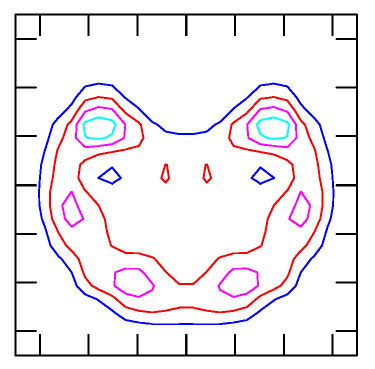}
\caption{Pair correlation functions for selected states of
seven electrons. Left panel shows the purely rotational state 
(ground state for $L=117$). The center panel shows a vibrational 
state (fourth excited state for $L=133$), which consists of vibrational modes
$\omega_5$ and $\omega_6$. The right panel shows 
a seven-fold ring as a high excited state (17th state for $L=133$).
}
\label{sevenfold}
\end{figure}

Finally, we would like to note that in the case of seven
particles the classical geometry of a seven-fold ring is also
stable. However, its energy is so much higher that it
does not have any contribution to the low energy spectrum
at any angular momenta. The situation is different to that
of six particles, where the ground state of a five-fold ring
(with a particle at the center) and the six-fold ring are very close
in energy and both contribute to the rotational 
spectrum\cite{maksym1996,manninen2001}. Nevertheless,
at higher energies the seven-fold ring also exists.
As an example, 
Fig. \ref{sevenfold} shows the pair correlation function of 
the $17^{\text{th}}$ state at $L=133$. Clearly,
it shows seven particles in a ring with an empty center.

\subsection{Gaussian interparticle interaction}

In order to study the effect of the range of the interparticle 
interaction to the localization of particles we used Gaussian
potential, Eq. (2). In this case we study a rather simple case of
four particles. The ground state classical configuration
of four particles is a square.
Prus {\it et al}\cite{prus2004} have developed a model that considers electrons
as Gaussian probability densities.
They showed that this model yields also to the formation of a Wigner
molecule. At this point we want to stress that our model is different from
that because here the particles are point like but the interaction
between the particles is Gaussian. 
Solving the eigenmodes is a bit more tricky
than it was with the Coulomb interaction because
the form of the interparticle interaction leads
into transcendent equations at some point.
As in the case of Coulombic interaction, knowing the modes is
important,  because the symmetry properties of the mode define all the
allowed combinations of angular momenta and vibrational modes.
In the three particle situation this point is reached
at the end of the calculations so that it is easy to keep
accounts for which energy value belongs to which mode.
The four particle problem, however, is somewhat more
troublesome because the stage of transcendent equations
comes before the diagonalization of the dynamical matrix.
This leads to difficulties of keeping track of
eigenmodes and eigenergies for different angular momenta.
We solved this problem by writing a program
that finds out the right mode for a certain eigenenergy
for every angular momentum separately.

Due to the symmetry, the vibrational modes are 
identical with the modes 
obtained in \cite{maksym1996,nikkarila2007} with Coulombic 
interparticle interaction, and consequently
their symmetry properties are known from the previous
reports \cite{maksym1996,nikkarila2007}.
In the rotating frame the four particles have
two low-energy modes.
In the case of Gaussian interaction these two modes
cross each other when the angular momentum is increased.
The angular momentum where this happens depends on the
width $\sigma$ of the Gaussian potential.
This as such is interesting, and if the classical model describes
the behavior of the quantum mechanical system correctly, the same
phenomenon should also occur in the quantum
mechanical spectrum. We will return to this detail later
after we have presented the results for quantum mechanical systems
and the corresponding classical systems. 

\begin{figure}[h]
\includegraphics[width=0.9\columnwidth]{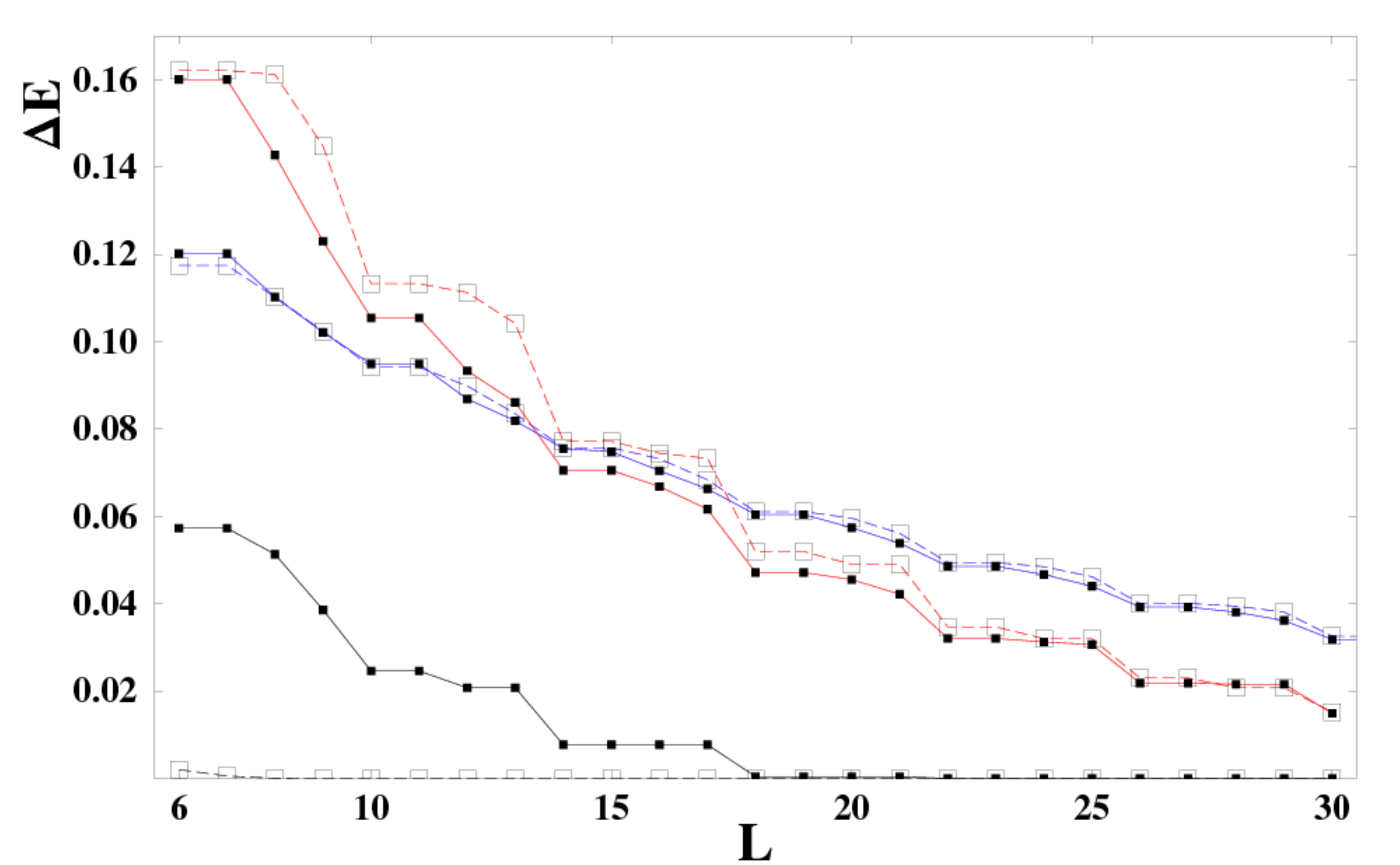}\\
\includegraphics[width=0.9\columnwidth]{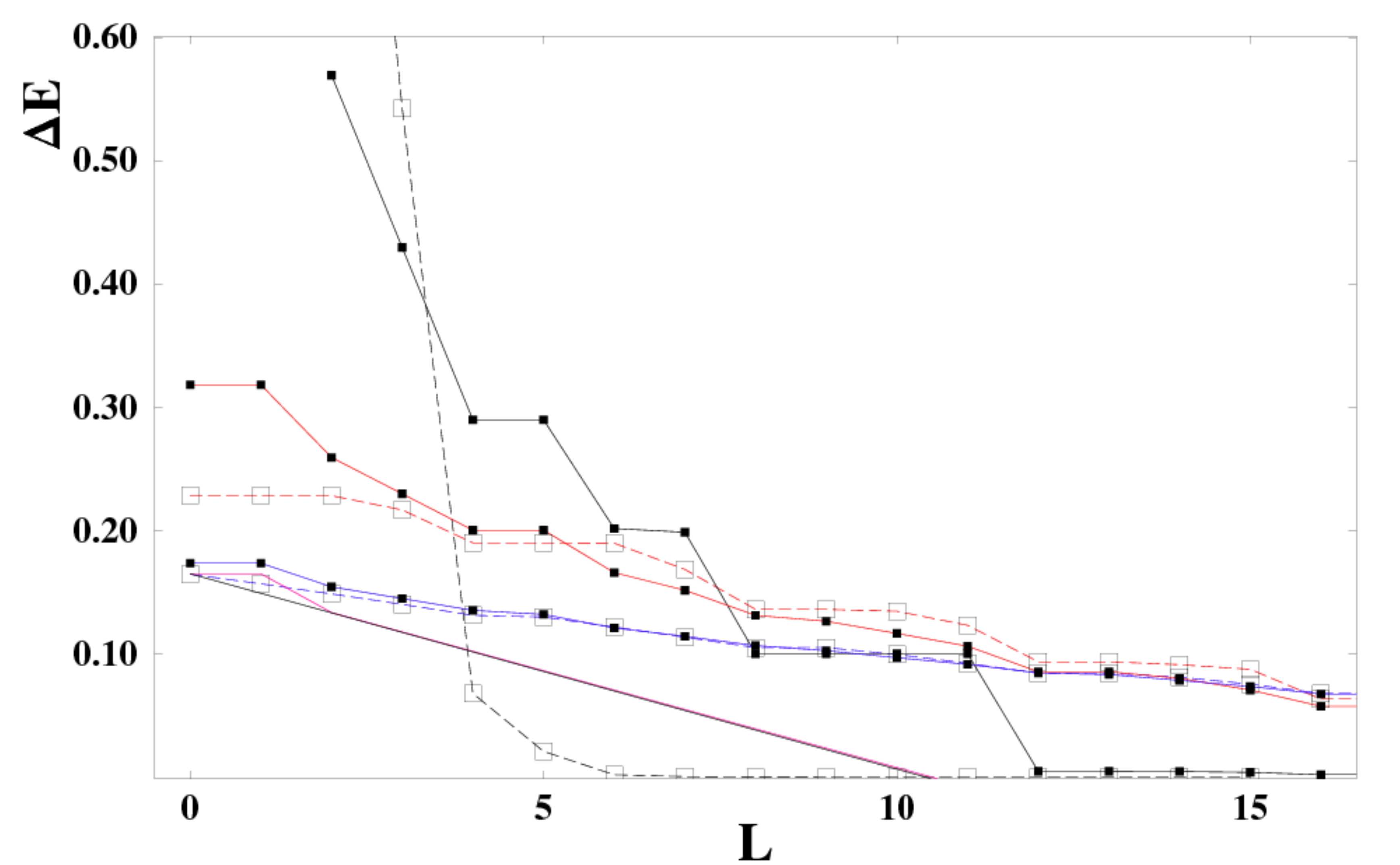}\\
\caption{Interaction energy for four fermions (uper panel) and 
four bosons (lower panel) as a function of the angular momentum,
calculated using different widths $\sigma$ of the Gaussian interaction.
Black dots connected with solid lines show the results 
of the exact diagonalization
in the LLL and the open circles connected with dashed lines
the results of the model Hamiltonian (8). 
Red curves, $\sigma=3.0$; blue 
curves, $\sigma=2.0$; black curves; $\sigma=0.5$.
The solid black and red lines (without points) show the 
exact and model Hamiltonian results for a harmonic interparticle
interaction.
}
\label{spectrum4}
\end{figure}

The quantum mechanical and classical energies for Gaussian repulsion
in the case of four fermions and bosons are represented in Fig.
\ref{spectrum4}
for three different values of $\sigma$: $0.5$, $2$ and $3$.
The results show that the behavior of the quantum mechanical systems approach
to the behavior of corresponding classical systems
when the Gaussian interparticle potential is broadened from a nearly 
contact potential
into a smoother potential. When $\sigma=3$, the quantum mechanical
spectrum is nearly identical with the corresponding classical
spectrum.
When $\sigma$ is decreased the classical and quantum mechanical 
results start to depart, although the similar structure survives
until very small values of $\sigma$.

Figure \ref{spectrum4}
also shows the results for a delta-function interaction
($\sigma=0$) and for a repulsive harmonic interaction
($v(r)=1/(\pi\sigma^2)(1-r^2/\sigma^2)$, with $\sigma=3$).
In the case of the delta function interaction the
interaction energy of the classical system is zero while
the quantum mechanical particles feel each other up to
Laughlin states where the contact interaction disappears,
i.e. up to $L=N(N-1)/2$ for fermions and $L=N(N-1)$ for bosons.
Beyond these angular momenta, the quantum mechanical and 
classical energies trivially agree. 
At this point it is interesting to note that in
quasi-one-dimensional quantum rings of fermions,
where the spectrum is dominated by particle localization at
all angular momenta, 
even a delta function interparticle interaction 
leads to vibrational states\cite{viefers2004}.

In the case of the repulsive harmonic interaction,
Eq. (5), the classical result approaches fast the 
quantum mechanical result when $\lambda$ increases,
as seen from Eqs. (6) and (10). Figure \ref{spectrum4}
show results for the value $\lambda$ corresponding to
the leading term of Gaussian interaction with $\sigma=3.0$.
In this case the classical and quantum mechanical results
follow the same line shown in the figures.
Since the classical approach gives correct results 
in the delta function limit and for the wide Gaussian limit
it is not surprising that it describes the quantum mechanical
spectra nearly exactly for all values of $\sigma$ 
when $L>N(N-1)$.

We have also made similar systematic study for three particles and
limited computations for seven particles using the Gaussian
interactions. The results are 
qualitatively similar to those presented here for four particles.

\begin{figure}[h]
\includegraphics[width=0.9\columnwidth]{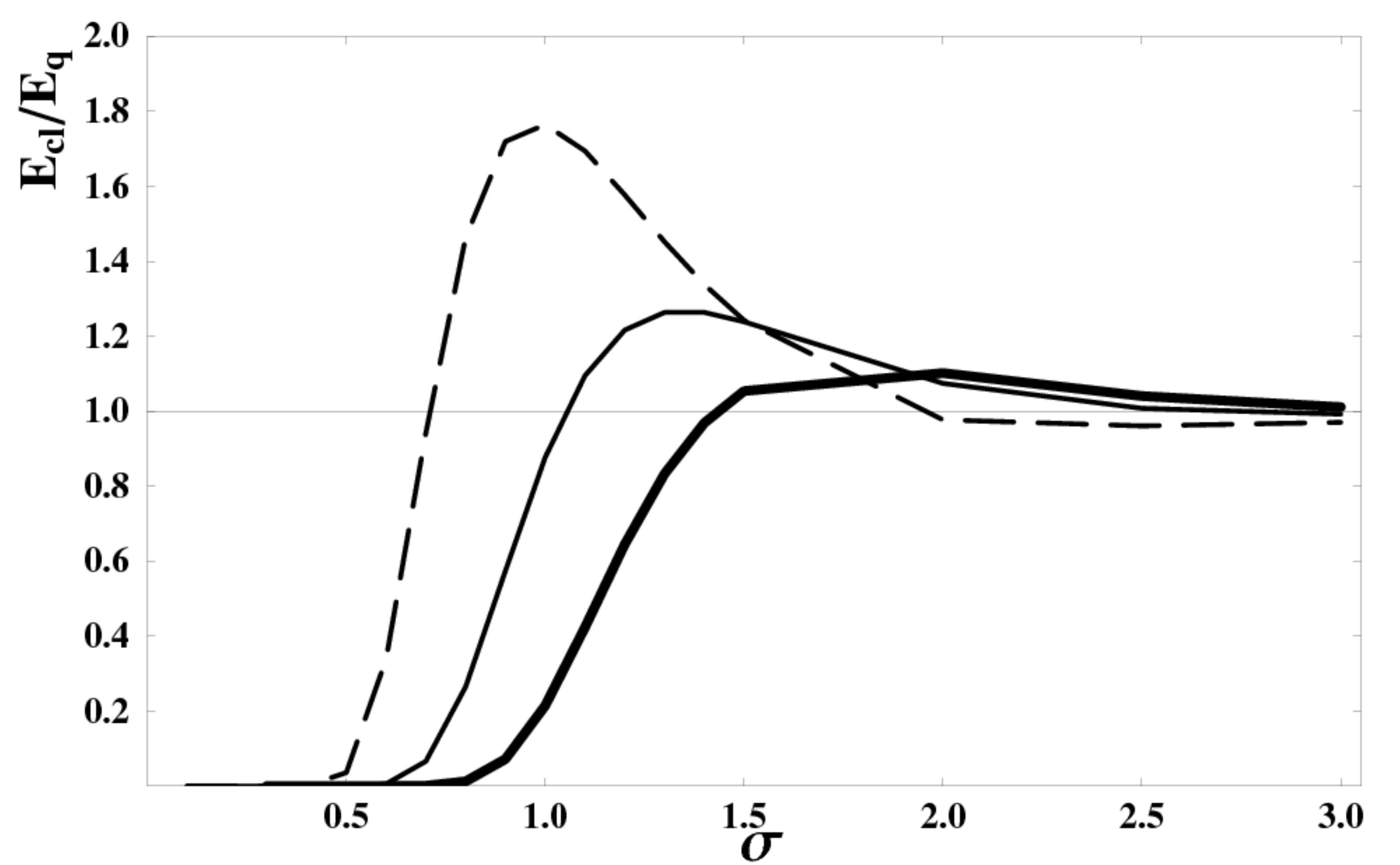}\\
\includegraphics[width=0.9\columnwidth]{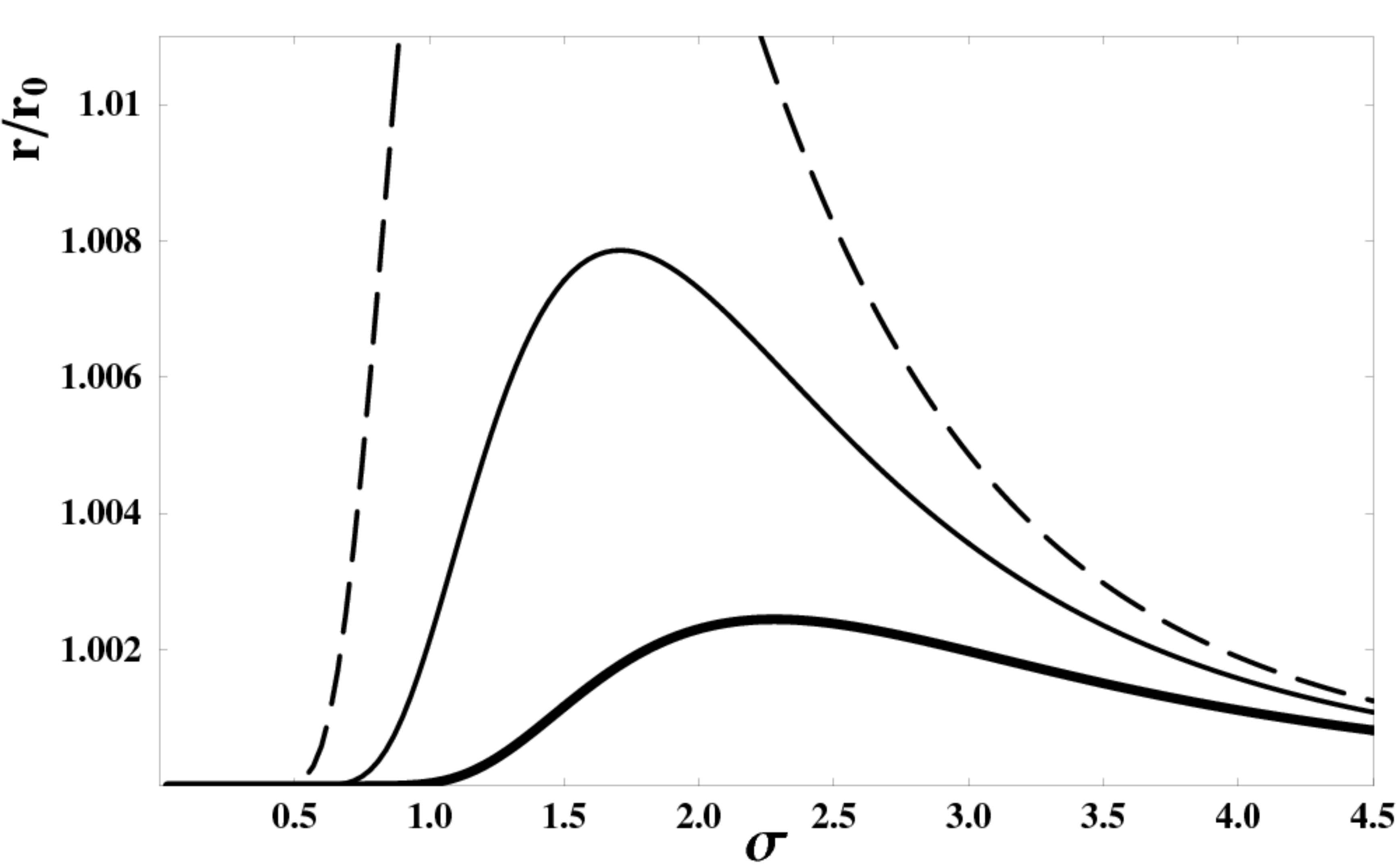}
\caption{Upper panel: Ratio of the ground state energy of the
classical model Hamiltonian and
the exact diagonalization energy
$E_{\text{cl}}/E_{\text{q}}$ as a function of the 
width of the Gaussian interaction $\sigma$.
Lower panel: Radius $r$ of the classical
equilibrium polygon (a square) as a function of $\sigma$. 
$r_0$ is the radius for $\sigma=0$). 
Dashed lines are for angular
momentum $L=6$, solid lines for $L=10$ and the thick solid
lines for $L=18$.
}
\label{E-b-sigma}
\end{figure}

The classical model seems to describe very well the quantum mechanical 
system especially at higher values of $\sigma$. In figure
\ref{E-b-sigma} we present a more detailed study of the effect
of $\sigma$ to the energy spectra. On the upper panel 
we present the ratio of classical and the quantum mechanical
ground state energies
as a function of $\sigma$ for a few fixed angular momenta.
On the other hand the results of the lower panel show
how the radius of the classical equilibrium polygon  (i.e a square)
depends on the value of $\sigma$.
The radii are normalized to one for every
angular momenta when $\sigma \approx 0$ in order to draw
the different radii on the same picture.
The behavior of the radii means that for a fixed
angular momentum the equilibrium configuration (i.e a square)
first expands as a function of $\sigma$ until it reaches its maximum
value, and after that starts to shrink.
The results on the two figures show a significantly similar behavior
of the two variables as a function of $\sigma$.
The explanation of such similarity is that classical
particles do not feel each others at all if the 
Gaussian potential is steep. When the potential 
broadens the potentials of different particles
start to overlap. Quantum mechanical particles on the other hand
feel each others even if the interparticle potential
is a point contact (i.e $\sigma \approx 0$).

\begin{figure}[h]
\includegraphics[width=0.9\columnwidth]{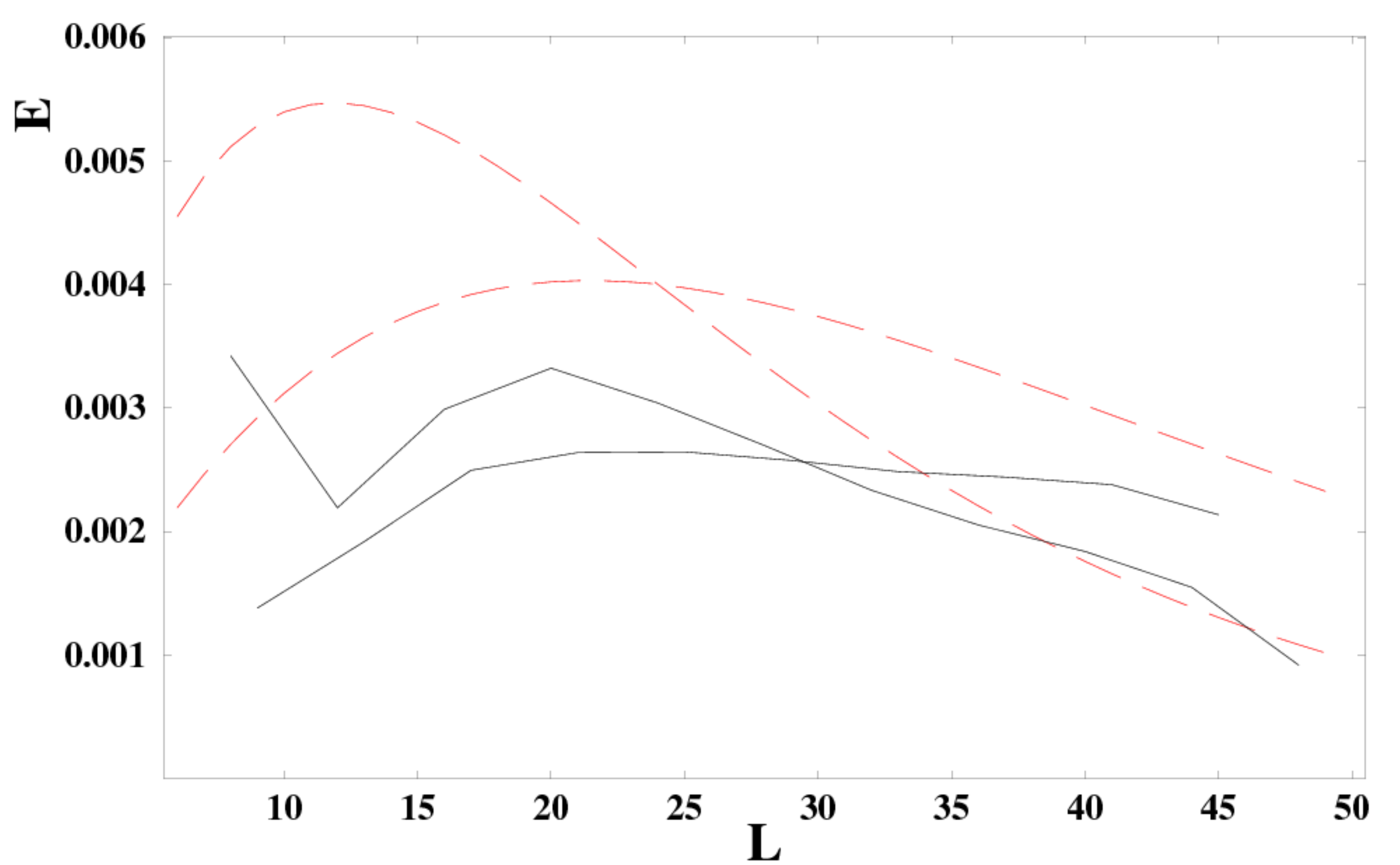}
\caption{Dependence of the two low-energy vibrational modes
on the total angular momentum in the case of four fermions
interacting with a Gaussian interaction with $\sigma=3.0.$
Dashed red lines show the classical results and 
solid black lines the quantum mechanical results
from the exact diagonalization.
}
\label{w3w4}
\end{figure}

We will now return to the crossing of the energies of 
the two lowest vibrational modes, 
when the angular momentum increases.
The symmetry dictates that one of these modes is 
the lowest energy state at angular momenta $L=4n$ in the case of 
fermions ($L=4n+2$ for bosons)
and the other mode at angular momenta $L=4n+1$
for fermions ($L=4n+3$ for bosons).
In order to get an estimate of the vibrational energy
from the exact diagonalization
we have to subtract from the total interaction energy 
the energy of the rigid rotation. 
This we have estimated by fitting a smooth function of 
angular momentum through the energies of 
the purely rotational states $L=6,10,14,\cdots$
for fermions.
The resulting vibrational energies for 
the Gaussian width $\sigma=3.0$
are plotted in Fig. \ref{w3w4}, together with the
classical vibrational energies, as a function of the 
angular momentum.
The results show that both classical and quantum mechanical
vibrational energies 
cross each other at about the same angular momentum.
This finding supports the previous notions
that the properties of the quantum mechanical
system of a few particles rotating in a harmonic confinement
can be explained with finest details by a model of
vibrating molecule of localized particles.

\begin{figure}[h]
\includegraphics[angle=-90,width=0.9\columnwidth]{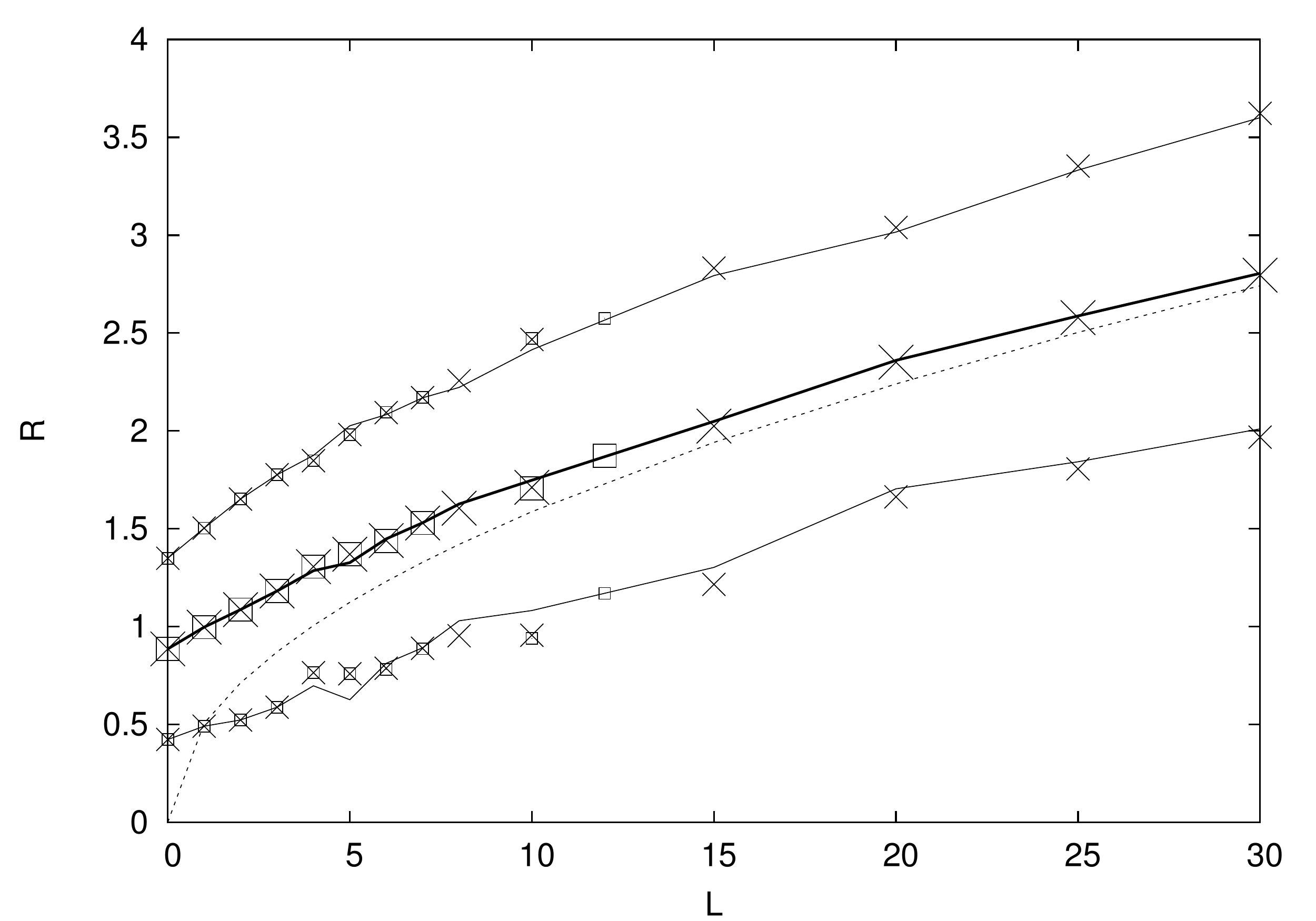}\\
\caption{Average radius of the ring of four bosons
as a function of the angular momentum $L$, calculated with
exact diagonalization in the LLL. The center
line and points show the radius $R$ and the lower and upper lines 
and points $R\pm W$, where $W$ is the variance of the radius.
Solid lines show results for the Gaussian interaction with 
$\sigma=3.0$, squares for $\sigma=0.05$ and crosses 
for the Coulomb interaction. Dotted line is the result for
classical particles with Gaussian interaction, $\sigma=3.0$.
}
\label{width}
\end{figure}

\begin{figure}[h]
\includegraphics[width=0.4\columnwidth]{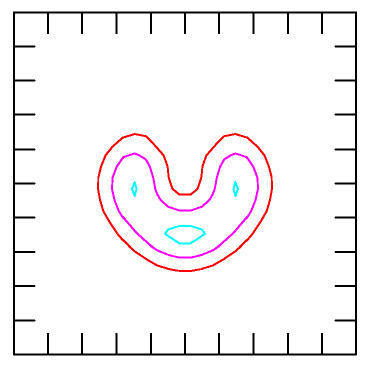}
\includegraphics[width=0.4\columnwidth]{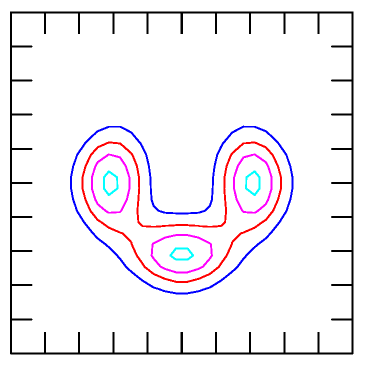}\\
\includegraphics[width=0.4\columnwidth]{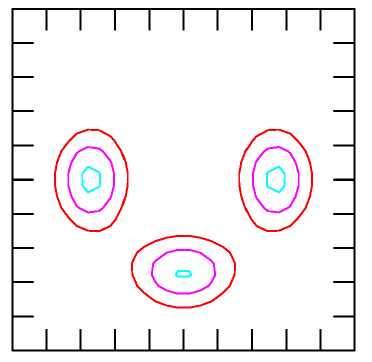}
\includegraphics[width=0.4\columnwidth]{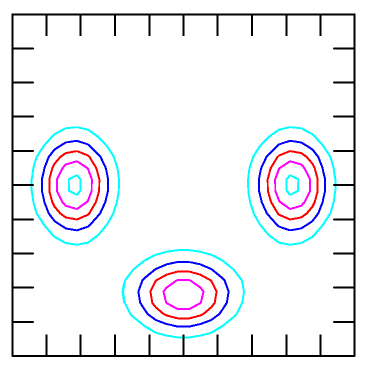}
\caption{Pair correlation functions of four fermions interacting with
the Gaussian interaction with $\sigma=3.0$.
Upper row shows results for $L=10$ (left)and $L=18$ (right)
and the lower row for $L=30$ (left) and $L=42$ (right).
The contour plots are in the same scale to demonstrate 
the expansion due to the rotation.
}
\label{pc4}
\end{figure}

Finally, we would like to quantify the degree of localization
in the quantum mechanical many-particle system. The symmetry
of the Hamiltonian dictates that the total particle density is 
circularly symmetric. However, for small number of particles they
localize in circles and consequently the radial density distribution
gives direct information of the degree of localization.
In the case of four electrons we have determined the
average radius $R$ of the ring of electrons and its variance $W$,
which describes the width of the localized electron state:
\be
R=\frac{1}{N}\int d{\bf r} r\rho(r),\quad 
W^2=\frac{1}{N}\int d{\bf r} (r-R)^2\rho(r),
\ee
where $\rho$ is the total electron density.
It turns out that for the quantum mechanical system both 
the radius and its variance are nearly independent of the shape
or strength of the interparticle interaction. The results for bosons
are shown in Fig. \ref{width} as a function of the angular momentum.
The radius reaches fast the classical value
and it is nearly independent of the interparticle interaction.
Similarly the width is independent of the interaction and also
nearly constant as a function of the angular momentum. 
The increased localization is mainly due to the increase of the 
radius with $L$ which makes the relative width to decrease.

The same result is seen qualitatively in Fig. \ref{pc4}
where the pair correlation function for four fermions with 
Gaussian interaction ($\sigma=3.0$) is shown for four different values
of the angular momentum. The pair correlation function demonstrates
that the angular localization increases with the same rate as
the radial localization.

\subsection{Rotation in a  3D harmonic confinement}

Above we have considered rotational spectra of interacting 
particles in a 2D harmonic potential. In the 3D case the 
classical particles interacting with repulsive interation
tend to form spherical shells\cite{manninen1986}.
However, when put into rotation, it is energetically favorable 
for the 3D structure to collapse into a 2D structure due to the 
increased momentum of inertia. For small number of particles 
this happens already at small angular momenta. 

As an example we have studied the case of seven particles interacting
with Coulomb interaction. The ground state geometry of a nonrotating
system is a pentagonal bipyramid, i.e. a five-fold ring in the
$x$-$y$-plane with two atoms at positions $\pm c\hat z$.
When this is put in rotation the five-fold ring will expand
while the particles in the $z$ axis will get closer to each other.
However, already at angular momentum $L=7$ the planar hexagon with
one particle at the center will have lower energy and 
at angular momentum $L=8$ the energy difference between
the 3D and 2D structures is much larger than the vibrational
energies of the 2D structure. 

Similar results are obtained for other small particle
numbers in 3D harmonic oscillator. Consequently, we can 
safely state that the region where the classical model
explains the spectrum for the 2D harmonic
oscillator the result would be the same had we done the computations
for the 3D confinement.

\section{Conclusions}

We have studied the rotation induced localization of particles,
in a two-dimensional circular harmonic trap.
Earlier studies have already shown that 
a long-range Coulomb interaction causes formation of 
Wigner molecules in quantum dots and that the quantum mechanical
many-particle spectra can be determined by 
canonical quantization of the rigid rotation and 
vibrational modes of the molecule\cite{maksym1996,nikkarila2007}.
The present calculations show that Coulomb interaction
also causes similar localization of bosonic particles.
For seven particles (fermions or bosons) the
low energy rotational spectrum 
is determined by vibrational modes of the molecule already
before the angular momentum corresponding to the filling
factor $\nu=1/2$ of the quantum Hall liquid is reached.

For mimicking atoms in a harmonic traps we have
used Gaussian interparticle interaction and studied 
the dependence of the localization on the range
of the interaction. It was observed that both in the 
short range (delta function) and long range (repulsive harmonic)
limit the classical model gives exactly the quantum
mechanical energies.
Consequently, for any range of Gaussian interaction the
classical model describes very accurately the quantum
mechanical many-particle energy spectrum at large enough
angular momenta.
However, the delta function interaction does not support
particle localization (particles do not feel each other).

We also estimated that in a three-dimensional harmonic 
confinement the rotating particles form very fast 
a two-dimensional configuration. For example, for seven
particles with Coulomb interaction this happens already 
at $L=7$, meaning that in the 3D confinement 
the low-energy rotational spectrum agrees with that
calculated for a 2D system.

In conclusion, independent of the form of the
repulsive interparticle interaction
or the type of the particles, the rotational spectrum of 
the particles confined in a harmonic trap can 
be estimated by quantizing the classical vibrational modes 
determined for localized particles in a rotating frame.

{\bf Acknowledgments}

We would like to thank Stephanie Reimann for valuable discussions. 
This work was supported by the Academy of Finland and by the 
Jenny and Antti Wihuri Foundation.

\end{document}